# A Distributed Software Architecture for Collaborative Teleoperation based on a VR Platform and Web Application Interoperability

Christophe Domingues, Samir Otmane, Fréderic Davesne and Malik Mallem
University of Evry – IBISC CNRS FRE 3190
40 Rue de Pelvoux, CE1455 Courcouronnes 91020 Evry Cédex, France
{Firstname.Surname}@ibisc.univ-evry.fr

## Abstract

*Augmented Reality and Virtual Reality can provide to a Human Operator (HO) a real help to complete complex tasks, such as robot teleoperation and cooperative teleassistance. Using appropriate augmentations, the HO can interact faster, safer and easier with the remote real world. In this paper, we present an extension of an existing distributed software and network architecture for collaborative teleoperation based on networked human-scaled mixed reality and mobile platform. The first teleoperation system was composed by a VR application and a Web application. However the 2 systems cannot be used together and it is impossible to control a distant robot simultaneously. Our goal is to update the teleoperation system to permit a heterogeneous collaborative teleoperation between the 2 platforms. An important feature of this interface is based on different Mobile platforms to control one or many robots.*

## 1. Introduction

Early Computer Assisted Teleoperation architectures provided to the human operator many kinds of assistances [1][2]. However diverse assistance strategies and information's concerning tasks are proposed and the robot commands where a function of operator actions and the adopted strategy (e.g Freezing some robot Degrees of freedom). The problem with this approach is that the current solutions are used in closed platforms and do not allow changes of functionalities [3].

Virtual Reality (VR) and Augmented Reality can benefit from teleoperation systems to improve human robot interaction. VR/AR has been used to solve time delay problems using predictive display concept where a virtual robot is surrounding the real robot on the video feedback [4][5][6][7].

In our laboratory, we have worked several years on human machine interaction model for teleoperation systems that led to the ARITI system. ARITI permits to control a 6 DoF robot with the use of Virtual Reality and Augmented Reality. Two teleoperation platforms exist. The first using a Virtual Reality/Augmented Reality semi immersive platform [8][9] and the second is an online version accessible through a Web browser [6].

Recently, collaboration functionalities have been integrated on the VR [9] platform and Web platform [10]. However, technical problems and utilization limitations appears that are described on this paper. The major problem is that the 2 platforms cannot be used at the same time. Consequently, collaboration is not possible between the 2 platforms. Collaborative teleoperation has been motivated by several situations: physical limitation of manipulator's workspace or grasping large objects. However as pointed in [11], one of the current challenges in robotics is the joint execution of tasks involving human beings and robots at the same workspace, also called telepresence.

With this work, our final goal is to permit collaborative teleoperation (users-robots) with multiple interfaces and multiples devices. So we want to make possible collaboration between VR/AR users, Mobile users and distant robots in a robot teaching application.

In the second section of the paper, the previous work of ARITI interface is presented with the different kinds of assistance. Finally, the third section is dedicated to the software and network architecture.

## 2. Previous Work

The ARITI project[1] or Augmented Reality Interface for Teleoperation via Internet has started several years ago. In fact, it is a Client/Server application, which allows visualizing and controlling a 6 DoF robot (a Fanuc Lr Mate) by using any remote computer. It is aimed at enhancing HO capabilities for achieving complex telerobotic operations.

In this part, we present the ARITI teleoperation system with its problems and limitations, where answers trying to solve the problems have permitted us to develop the work presented in this paper.

In order to help the HO to control the robot and achieve a complex teleoperation task, an interactive assistance is given to him, which is a Virtual Fixtures (VF) intervention in the operation area. These VF appear and disappear as the robot's peg comes closer or away from the objects to be manipulated [12]. This type of virtual fixtures appears according to the kind of tasks that had already been achieved.

---
[1] Web site for manipulating robots http://ariti.ibisc.univ-evry.fr

Figure 1 presents ARITI versions' history. The first application created, is ARITI Web mono user. After that, we wanted to integrate collaborative functions on the application and in parallel we wanted to develop the VR platform to improve user experience using 3D vision and natural interaction.

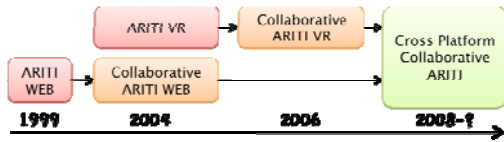

Figure 1: The ARITI versions history and the final goal of our work.

### 2.1. ARITI Web

ARITI Web is based on a Java Applet where the user interface is divided in 4 parts. First, the augmented video feedback of the distance robot with the virtual robot superposed on. Second, there are 2 virtual views of the virtual robot. Finally, the last part presents the way to control the robot by clicking adequate buttons. At the beginning, collaboration was not possible therefore, the application has been updated to integrate these concepts where users form a group and each user is affected to a mission that they perform independently.

### 2.2. ARITI on VR Platform

At the beginning, ARITI VR (see figure 2) was used with a 6 DoF with only mono user support, where the collaborative functions have been integrated later.

We have tested a collaborative function with another VR platform [9]. A Web interface was used to display the videoconference of the 2 distant operators while the second platform used the SPIDAR as an interaction device to control a virtual robot. The first platform used a Flystick as interaction device to control the virtual robot but this virtual robot was superposed on the real robot by using a video feedback.

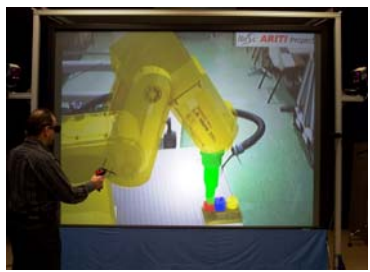

Figure 2: The operator controls the virtual phantom robot to take virtual objects. After the validation, the distant real robot will follow the phantom and take the real objects

### 2.3. Limitations and Technical Problems

The major limitation of the actual system is that the 2 versions are not compatible together. To use the Web version, we must stop the VR Platform. Our goal is to have a global teleoperation system that permits to teleoperate a robot with the desired platform.

The video stream of the distant robot can only be diffused for one user only because the protocol used to diffuse the stream is a homemade protocol. Consequently, when the second user connects to the system, he can only view the virtual robot.

The VR platform application is realized in C++ with the use of OpenGL library. Moreover, the interaction part is limited to the devices available on our VR platform (ART tracking, SPIDAR).

The Web application is realized in JAVA in an applet format. Collaboration is available but users must form a group and selected pre defined missions, as they cannot free control the robot together.

## 3. The new teleoperation system

The proposed teleoperation system tries to remedy the limitations and problems. In this part we present the proposed corrections and the technical solutions. Our intention is to support multiple interfaces and multiple devices.

### 3.1. Solving problems and limitations

We have integrated the Virtools solution to develop VR applications discussed in the previous section. Virtools is an application to easily develop VR applications for commercial or educational activities. Virtools natively integrates VRPN, an open source solution to use a multiple VR devices. The first version of ARITI Web uses Java Standard Applet. This solution needs the Virtual Machine installed on the client. Using the J2EE based on a Tomcat server permits to deport and execute the code on the server and not in the client. However, we could choose PHP or ASP as the portability is easier (J2SE -> JEE) and JAVA offering more possibilities due to the existence of JAVA 3D JAVA VLC (JVLC) libraries.

### 3.2. The new network architecture

To remedy the limitation of the non-use of the 2 platforms together, we have added the multi user server. This server permits to link VR and Mobile applications. Figure 3 shows the network architecture with the integration of the multi user server. This server is connected to a MySQL database storing position and orientation of all shareable objects but also the position and orientation of the users' virtual robots. The multi user server is connected to the robot server to transmit the robot commands from the users.

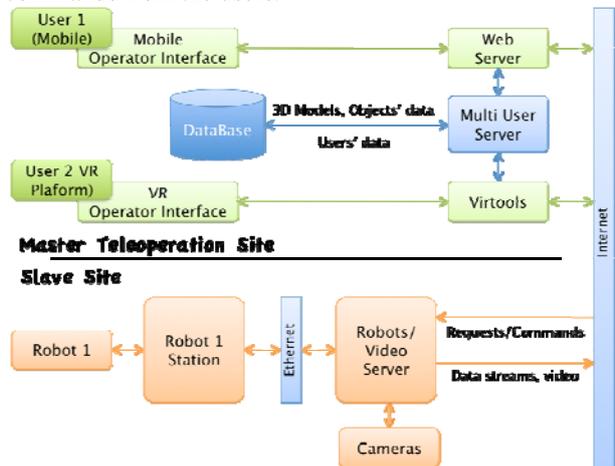

Figure 3: The network architecture for the new ARITI teleoperation system. The user can indifferent used a Web or VR application and can do collaborative teleoperation with others Web or VR users.

For the Video component, we have used Windows Media Services based on Microsoft Windows Server 2003 Enterprise Edition on the relay server. The relay server receives streams from different servers coder and diffuses (Multicast or Unicast) to the connected clients. The encoder servers used Microsoft Windows Media Coder 9. This solution has been chosen instead of ffmpeg streaming solution because streaming with ffmpeg does not work correctly with Virtools.

The media coder can encode video and audio from different sources only the sources are DxShow compatible (e.g. Webcams or Firewire cameras). We have used two types of cameras: Logitech Quickcam Pro 4000 and PixeLink Firewire cameras.

### 3.3. The new software architecture

With this work, we wanted to integrate on the software architecture a system that will permit the personalization of the application. To achieve this objective, we have considered all the functionalities as modules where each module is designed with the same conventions and is available in 2 versions (Mobile and classic).

A module has 2 ways of working (Safe or classic mode) and communicates with the core. There are 2 cores version, the mobile and the VR Core. The VR core is implemented in Virtools scripting language while the second is implemented in JAVA. Despite of the existence of 2 core versions, each core has the same structure, functions, inputs and outputs.

The core sends to a module synchronization signals as "LOAD" or "UNLOAD". These signals are used to load the module or unload the module when the application is running. Figure 4 illustrates the communication between the core and different modules. At the beginning, modules are loaded by the module loader (component of the core). When the load is complete, the module gives a state report to the core that permits to determine the working mode (Safe or classic). A safe mode permits to deactivate some possibilities of the module by sending an adequate signal ("SAFE"). The core also sends with the signal an argument. This argument is a degree of degradation.

For example, the user has selected 5 cameras to get 5 point of view of the distant robot but the latency is too high. So this core will automatically use the maximum number of cameras possible to have a good latency in order to preserve a good experience for the user, the module receives "SAFE 4" with the argument "4" for 4 cameras. Figure 4 shows the software architecture with the use of 3 modules.

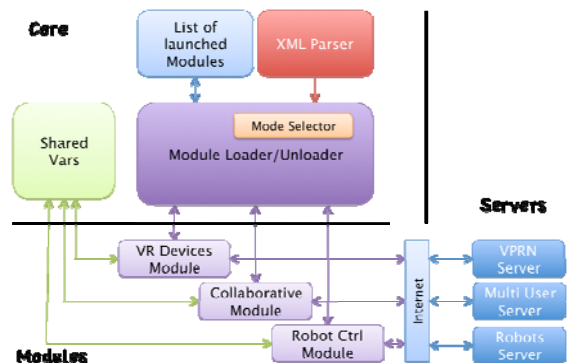

Figure 4: The communication between the VR/Mobile core, the modules and adequate servers. The cores have the same architecture just modules are different. VRPN server permits to retrieve data from VR devices.

The main advantage of this architecture is that the user can dynamically add modules in the application without stopping it. For example, during a teleoperation task the user has initially chosen to free control the robot, however he needs to program a trajectory and stored it in the robot. In this case, the system is available to add the trajectory module when the application is running. The prototyping system uses the web server of the mobile applications.

The user will use the Prototyping System to select all the elements he wants to use in the teleoperation application. In fact, a module is a package, which contains the module and an XML description of the module, which is containing methods and arguments. A final XML file is created which is containing all modules description and the user option. Figure 5 illustrates the Prototyping System. A list of modules is displayed. This list is stored on a MySQL database.

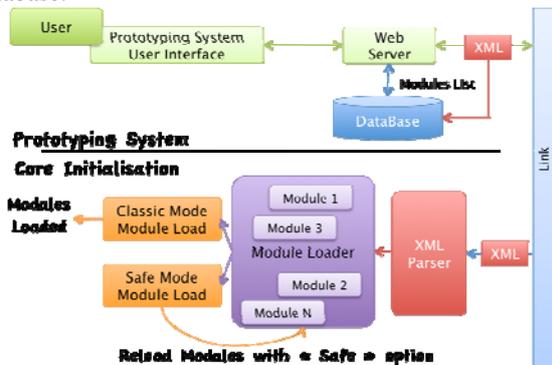

Figure 5: The XML file is an input parameter of the Core. At the end, the result is a VR or a Web application with adequate loaded modules. Modules can be loaded in safe mode.

## 4. Conclusion

We presented a new collaborative Man Machine Interface (MMI) allowing high-level teleworking activities and collaborative robot teleoperation using Augmented and Virtual Reality technologies. We started this work by identifying problems and limitations on actual systems. In fact, we wanted to integrate collaborative robot-teaching aspects with the support of multi interfaces and multi devices. Consequently, we have built our architecture with this constraint. One of the major problems encountered was the video streaming to permit a synchronous diffusion of the distant robot video stream. The second condition is to permit to quickly and easily add/delete/update modules on the system. To answer these conditions we have proposed a Prototyping System, which permit to personalize an application by clicking on desired modules. PS permits to choose between a VR application compatible with the platform and a Web application usable on a Web Browser or any mobile device. We have tested the system with an iPhone, a Web browser and a VR application but no quantitative results have been gathered. An example of teleworking application is presented as a teleoperation of the remote 6 DoF Fanuc robot for robot teaching in a mixed environment. An important feature of this application is based on the use of different technological heterogeneous platforms to control the robot such TCP/IP. Future work will focus on many aspects as:

- Completing the transformation of the old ARITI system by developing missing modules;
- Performing evaluations on the interaction aspects (Mobile versus VR; Virtual versus Real, VR versus VR) with the use of an evaluation system [13] to improve the system;
- A new model of Virtual Fixtures as it exists on ARITI but for a collaborative case where users want to grasp the same object at the moment.